\documentclass[preprint,showpacs,showkeys,superscriptaddress,amssymb,numbers,sort&compress]{elsarticle}

\date{\today}

\usepackage{graphicx,amssymb,epstopdf,tabularx}

\makeatletter
\def\fnum@figure#1{\figurename\nobreakspace\thefigure
.}
\def\fnum@table#1{\tablename\nobreakspace\thetable
.} \makeatother

\begin{document}

\title{Isovector channel of quark-meson-coupling model and its effect on symmetry energy}

\author[rvt]{X.B. Wang}

\author[focal]{C. Qi}

\author[rvt,els]{F.R. Xu\corref{cor1}}
\ead{frxu@pku.edu.cn}

\cortext[cor1]{Corresponding author}

\address[rvt]{School of Physics, and State Key Laboratory of Nuclear
Physics and Technology, Peking University, Beijing 100871, China}
\address[focal]{KTH (Royal Institute of Technology), Alba Nova
University Center, SE-10691 Stockholm, Sweden}
\address[els]{Center for Theoretical Nuclear Physics, National
Laboratory for Heavy Ion Physics, Lanzhou 730000, China}

\begin{abstract}

The non-relativistic approximation of the quark-meson-coupling model
has been discussed and compared with the Skyrme-Hartree-Fock model
which includes spin exchanges. Calculations show that the
spin-exchange interaction has important effect on the descriptions
of finite nuclei and nuclear matter through the Fock exchange. Also
in the quark-meson-coupling model, it is the Fock exchange that
leads to a nonlinear density-dependent isovector channel and changes
the density-dependent behavior of the symmetry energy.

\end{abstract}


\maketitle

\section{Introduction}\label{intro}\noindent

Both relativistic and non-relativistic energy density functional
models have been developed to describe the many-body system of
atomic nucleus~\cite{Benderreview}. According to the Kohn-Sham
theorem~\cite{Kohnshamtheory}, models leading to the same energy
functional are equivalent as far as physical results are
concerned~\cite{lowdensityexpansion}. Indeed, systematical
calculations have shown that both relativistic and non-relativistic
models give quite similar results for the ground-state properties
and low-energy dynamics of stable nuclei~\cite{Stonereview}.
However, significant differences are found when calculations go to
the extreme cases of high isospin and high
density~\cite{Benderreview, Libaoan}. The origin of these
differences is still not clear.

In the non-relativistic approach, the Skyrme-Hartree-Fock (SHF)
model has been extensively employed in the investigation of nuclear
structure~\cite{Benderreview,Stonereview} since the first work of
Vautherin and Brink~\cite{Skyrmefirst}. The Skyrme potential
contains ten parameters, but only six of the parameters have been
well determined~\cite{six}. This leads to many sets of parameters
with different emphases on the descriptions of nuclear
properties~\cite{stone, Reinhardreview}. It was tested by Stone {\it
et al.}~\cite{stone} that about one third of existing Skyrme
parameter sets give the monotonic increase of the symmetry energy
with density while other two third give the decrease of the symmetry
energy after certain supra-saturation density. It should be
mentioned that predictions for the density dependence of the
symmetry energy depend strongly on models and parameters
employed~\cite{Libaoan,neutronskin,softsymmetrycurve}. As examples,
the
Variational-Many-Body~\cite{Vmbsymmetry1,Vmbsymmetry2,Vmbsymmetry3}
and Dirac-Brueckner-Hartree-Fock~\cite{dbhfsymmetry} theories
predict that the symmetry energy turns to decrease after certain
supra-saturation density, while RMF~\cite{Libaoansymmetry,Liu02} and
Brueckner-Hartree-Fock~\cite{bhfsymmetry} models present that the
symmetry energy increases monotonically with increasing density.
Therefore, it is encouraging to investigate possible physics origins
for the different behaviors of the symmetry energy at high
densities~\cite{mbtsang1,mbtsang2,Fuchs,Libaoan} which may shed
light on our understanding of nuclear force.

Based on the mean-field approximation but including the response of
the quark structure of the nucleon in the nuclear environment, the
relativistic quark-meson-coupling model (QMC) was proposed by
Guichon \textit{et al.}~\cite{qmcorigin,qmcfinite}. In
Ref.~\cite{qmcnpa1} an interesting similarity was found by comparing
the energy density functionals of the SHF model and the
non-relativistic approximation of QMC models, providing an
understanding of the physical origin of the Skyrme force within the
relativistic approach. With the restriction of $N=Z$, the
authors~\cite{qmcnpa1} obtained a set of parameters of the Skyrme
force that gives reasonable descriptions of nuclear matter and
finite nuclei. The work suggests that investigations regarding the
physics of effective nucleon-nucleon interactions remain
interesting. In Ref.~\cite{qmcnpa1}, the Skyrme spin-exchange terms
that involve the $x_1$, $x_2$, and $x_3$ parameters are neglected.
The present discussion will include the spin-exchange terms. It will
be shown that these terms have important effects on nuclear
properties at high isospin and high density, particularly on the
density dependence of the symmetry energy.

\section{Energy density functionals of QMC and Skyrme Hartree-Fock models}

The QMC model gives a relativistic quark-level description of
nuclear matter~\cite{qmcorigin,qmcfinite}. In the model, a baryon in
nuclear medium is assumed to be a non-overlapping static spherical
bag in which quarks are coupled to meson fields in the mean-field
approximation~\cite{qmcorigin,qmcfinite,qmcnpa1,QMCreview}. For
modeling, one can choose an isoscalar-scalar field ($\sigma$) for
the medium-range attraction, an isoscalar-vector field ($\omega$)
for the short-range repulsion, and an isovector-vetor field ($\rho$)
for isospin channel~\cite{qmcnpa1}. Detailed discussions on the
Hartree-Fock (HF) approximation of the QMC model can be found in
Refs.~\cite{qmcfockcqm,qmcfockcbm,qmcnpa1,qmcnpa3}.

As has been given in Ref.~\cite{qmcnpa1}, for finite nuclei, the
non-relativistic HF approximation of the QMC density functional can
be written as follows
\begin{equation}\label{qmc}
{\cal E}^{\rm QMC}=\rho M+\frac{\tau}{2M}+{\cal H}_{0}+{\cal
H}_{3}+{\cal H}_{\rm eff}+{\cal H}_{\rm fin}+{\cal H}_{\rm so},
\end{equation}
with
\begin{eqnarray}\label{qmc03}
\nonumber{\cal H}_{0}+{\cal H}_{3}&=&
\rho^2\left[\frac{-3{G_\rho}}{32}+\frac{G_\sigma}{8\left(1+d\rho{G_\sigma}\right)^3}
-\frac{G_\sigma}{2\left(1+d\rho{G_\sigma}\right)}
+\frac{3{G_\omega}}8\right] \\
&&+ \left(\rho_n-\rho_p\right)^2\left[\frac{5{G_\rho}}{32}+
\frac{G_\sigma}{8\left(1+d\rho{G_\sigma}\right)^3}-\frac{{G_\omega}}8\right],
\end{eqnarray}
\begin{eqnarray}\label{QMC_eff}
\nonumber{\cal H}_{\rm eff}
=&&\left(\frac{G_\rho}{4m_\rho^2}+\frac{G_\sigma}{2M^2}\right)\rho\tau\\&&+\left(-\frac{G_\rho}{8m_\rho^2}-\frac{G_\sigma}{2m_\sigma^2}+\frac{G_\omega}{2m_\omega^2}
-\frac{G_\sigma}{4M^2}\right)(\rho_n\tau_n+\rho_p\tau_p),
\end{eqnarray}
\begin{eqnarray}\label{QMC_fin}
\nonumber{\cal H}_{\rm fin}&= &\left(\frac{-3G_\rho}{16m_\rho^2}-
\frac{G_\sigma}{2m_\sigma^2}+\frac{G_\omega}{2m_\omega^2}-
\frac{G_\sigma}{4M^2}\right)\rho\nabla^2\rho\\
&&+\left(\frac{9G_\rho}{32m_\rho^2}+\frac{G_\sigma}{8m_\sigma^2}-
\frac{G_\omega}{8m_\omega^2}+\frac{G_\sigma}{8M^2}\right)
(\rho_n\nabla^2\rho_n+\rho_p\nabla^2\rho_p),
\end{eqnarray}
and
\begin{eqnarray}\label{QMC_so}
\nonumber{\cal H}_{\rm so}&=&
\left(\frac{-G_\sigma}{4M^2}+\frac{G_\omega\left(1-2\mu_s\right)}
{4M^2}\right)\rho\nabla\cdot J\\\nonumber
&&+\left(\frac{-G_\sigma}{8M^2}+\frac{G_\omega\left(1-4\mu_s\right)}{8M^2}-
\frac{3G_\rho\left(-1+2\mu_v\right)}{32M^2}\right)\\&&\times(\rho_n\nabla\cdot
J_n+\rho_p\nabla\cdot J_p),
\end{eqnarray}
where $M$ is for the mass of the nucleon, $\tau=\tau_n+\tau_p$ for
kinetic energy density, and $\rho=\rho_n+\rho_p$ for density with
$n$ and $p$ denoting the neutron and proton, respectively.
$G_\sigma=g^2_\sigma/m^2_\sigma$, $G_\omega=g^2_\omega/m^2_\omega$
and $G_\rho=g^2_\rho/m^2_\rho$ are defined with $g_\sigma$,
$g_\omega$, $g_\rho$ for coupling constants and $m_\sigma$,
$m_\omega$, $m_\rho$ for the masses of the mesons. The parameter $d$
is the scalar polarizability of the nucleon~\cite{qmcnpa1}. In the
numerical calculations of the present work, we take $d=0.150$ fm
from Ref.~\cite{qmcnpa1}. $\mu_s$ and $\mu_v$ are isoscalar and
isovector magnetic moments of the nucleon, respectively.
$\mu_s=\mu_p+\mu_n$ and $\mu_v=\mu_p-\mu_n$ ($\mu_p=2.79\mu_{\rm
N}$, $\mu_n=-1.91\mu_{\rm N}$). $J_n$ and $J_p$ are the spin density
of the neutron and proton, respectively, with the  total spin
density of $J=J_n+J_p$.

In Refs.~\cite{qmcnpa1,qmcnpa2} it was shown that the six elementary
parameters ($t_0$, $t_1$, $t_2$, $t_3$, $x_0$ and $W_0$) of the
Skyrme force can be understood in the framework of the QMC model. In
the present work, we study in a more detailed energy density
functional of the SHF model with spin-exchange parameters ($x_1$,
$x_2$, $x_3$) which can produce effects through the Fock exchange.
We can have a general form of the Skyrme energy density functional
as~\cite{SHFeq},
\begin{eqnarray}\label{sk03}
\nonumber{\cal H}_{0}+{\cal H}_{3}=&
&\rho^2\left[\frac{{t_3}\rho^\alpha}{16}+\frac{3{t_0}}8\right] \\&&+
\left(\rho_n-\rho_p\right)^2
\left[-\frac{{t_0}\left(2{x_0}+1\right)}8
-\frac{\left({2x_3}+1\right){t_3}\rho^\alpha}{48}\right],
\end{eqnarray}
\begin{eqnarray}\label{sk_eff}
\nonumber{\cal H}_{\rm eff}=&&
\frac{1}{8}\left[\left(2+x_1\right)t_1+\left(2+x_2\right)t_2\right]\rho\tau\\&&+
\frac{1}{8}\left[t_2\left(2x_2+1\right)-t_1\left(2x_1+1\right)\right]
\left(\rho_n\tau_n+\rho_p\tau_p\right),
\end{eqnarray}
\begin{eqnarray}\label{sk_fin}
\nonumber{\cal H}_{\rm
fin}=&&-\frac{1}{32}\left[3t_1\left(2+x_1\right)-
t_2\left(2+x_2\right)\right]\rho\nabla^2\rho\\
&&+\frac{1}{32}\left[3t_1\left(2x_1+1\right)+t_2\left(2x_2+1\right)\right]
\left[\rho_n\nabla^2\rho_n+\rho_p\nabla^2\rho_p\right],
\end{eqnarray}
and
\begin{eqnarray}\label{sk_so}
{\cal H}_{\rm so}=-\frac{1}2W_0\left(\rho\nabla \cdot J+\rho_n\nabla
\cdot J_n+\rho_p\nabla \cdot J_p\right).
\end{eqnarray}

The QMC and SHF models have the same density-dependent forms in the
effective-mass term ${\cal H}_{\rm eff}$ and the finite-range term
${\cal H}_{\rm fin}$. By comparing ${\cal H}_{\rm eff}$ and ${\cal
H}_{\rm fin}$, we obtain the following relations,
\begin{eqnarray}\label{t1}
t_1=\frac{G_\rho}{2m_\rho^2}+\frac{3G_\sigma}{2M^2}-
\frac{2G_\omega}{m_\omega^2}+\frac{2G_\sigma}{m_\sigma^2},
\end{eqnarray}
\begin{eqnarray}\label{x1}
x_1=\frac{2\frac{G_\rho}{m_\rho^2}}{\frac{G_\rho}{m_\rho^2}+
\frac{3G_\sigma}{M^2}-\frac{4G_\omega}{m_\omega^2}+\frac{4G_\sigma}
{m_\sigma^2}},
\end{eqnarray}
\begin{eqnarray}\label{t2}
t_2=-\frac{G_\rho}{2m_\rho^2}+\frac{5G_\sigma}{6M^2}+
\frac{2G_\omega}{m_\omega^2}-\frac{2G_\sigma}{m_\sigma^2},
\end{eqnarray}
and
\begin{eqnarray}\label{x2}
x_2=\frac{\frac{G_\rho}{m_\rho^2}-\frac{2G_\sigma}{3M^2}}
{-\frac{G_\rho}{2m_\rho^2}+\frac{5G_\sigma}{6M^2}+\frac{2G_\omega}{m_\omega^2}-\frac{2G_\sigma}{m_\sigma^2}}.
\end{eqnarray}

Comparing the energy density functionals of the QMC and SHF models,
we find that main differences in density dependence appear in the
terms which involve the parameters $t_3$ and $x_3$ in the zero-range
term ${\cal H}_{0}$ and the density-dependent term ${\cal H}_{3}$.
For a qualitative study, we expand the ${\cal H}_{0}+{\cal H}_{3}$
term around the saturation density in respect that the density range
of finite nuclei is around the saturation density. For the
coefficient of the $\rho^2$ term in the corresponding ${\cal
H}_{0}+{\cal H}_{3}$ term, the first and second orders of the
expansions give
\begin{eqnarray}\label{skqmc1}
\frac{-3{G_\rho}}{32}+\frac{G_\sigma}{8(1+d\rho_0{G_\sigma})^3}-
\frac{G_\sigma}{2(1+d\rho_0{G_\sigma})}+\frac{3{G_\omega}}8
=\frac{3{t_0}}8+\frac{{t_3}\rho_0^\alpha}{16},
\end{eqnarray}
and
\begin{eqnarray}\label{skqmc2}
-\frac{G_\sigma}8\frac{3d\rho_0{G_\sigma}}{\left(1+d\rho_0{G_\sigma}\right)^4}
+\frac{G_\sigma}2\frac{d\rho_0{G_\sigma}}{\left(1+d\rho_0{G_\sigma}\right)^2}
=\frac{t_3}{16}\alpha{\rho_0}^{\alpha},
\end{eqnarray}
where $\rho_0$ is the saturation density. Similarly, the expansions
of the coefficients of the $(\rho_n-\rho_p)^2$ term give the
following relations,
\begin{eqnarray}\label{skqmc3}
\frac{5{G_\rho}}{32}+\frac{G_\sigma}{8(1+d\rho_0{G_\sigma})^3}-
\frac{{G_\omega}}8
=-\frac{\left({2x_3}+1\right){t_3}\rho_0^\alpha}{48}
-\frac{{t_0}\left(2{x_0}+1\right)}8,
\end{eqnarray}
and
\begin{eqnarray}\label{skqmc4}
\frac{G_\sigma}8\frac{3d\rho_0{G_\sigma}}{\left(1+d\rho_0{G_\sigma}\right)^4}
=\frac{\left({2x_3}+1\right)}{48}{t_3}\alpha {\rho_0}^\alpha.
\end{eqnarray}

\begin{figure}
\hspace{-0.8  cm}\includegraphics[scale=0.5]{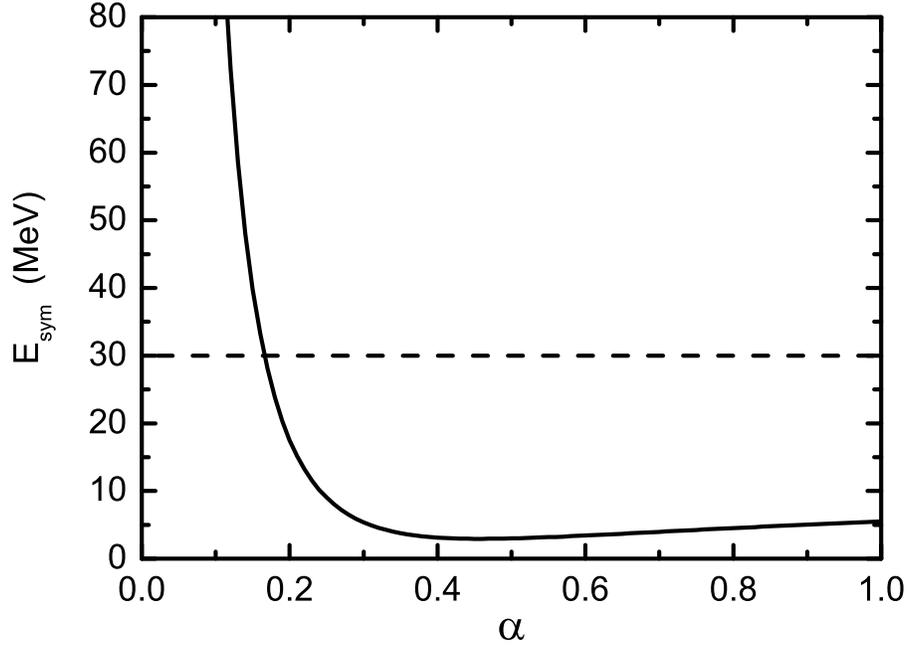}
\caption{The symmetry energy for nuclear matter at the saturation
density as a function of the $\alpha$ parameter, calculated using
the SHF model with the parameters obtained through the QMC model.
The dashed line indicates the commonly-adopted symmetry energy of
$a_{\rm s}=30$ MeV.}\label{figpowervalue}\end{figure}

From Eqs.~(\ref{skqmc2}) and (\ref{skqmc4}), the parameter $x_3$ can
be obtained immediately. The parameters $x_0$, $t_0$ and $t_3$ are
related to the parameter $\alpha$. We can use the symmetry energy of
the nuclear matter at the saturation density to determine the
parameter $\alpha$. Fig.~\ref{figpowervalue} plots the variation of
the symmetry energy with changing the parameter $\alpha$. At each
given $\alpha$ value, the parameters $x_0$, $t_0$ and $t_3$ are
obtained using Eqs.~(\ref{skqmc1})-(\ref{skqmc3}), and the SHF
symmetry energy~\cite{SHFeq} is calculated with these parameters
including those given by Eqs.~(\ref{t1})-(\ref{x2}). The QMC
parameters are taken from Ref.~\cite{qmcnpa1}. It is well accepted
that the symmetry energy of the nuclear matter at the saturation
density is about 30 MeV, which leads to an $\alpha$ value of about
$1/6$, that is consistent with the usual value in the SHF model, see
Fig.~\ref{figpowervalue}. With the $\alpha$ value and the QMC
parameters we can determine the $x_0$, $t_0$ and $t_3$ using
Eqs.~(\ref{skqmc1})-(\ref{skqmc3}).

For completeness, we take the $W_0$ parameter corresponding to the
spin-orbit term ${\cal H}_{\rm so}$ as Ref.~\cite{qmcnpa1}
\begin{equation}\label{W_0}
W_0=\frac{1}{12M^2}\left[5G_\sigma-5G_\omega\left(1-2\mu_{s}\right)-
\frac{3}{4}G_\rho\left(1-2\mu_{v}\right)\right].
\end{equation}

With the QMC parameters at $m_{\sigma}=700$ MeV~\cite{qmcnpa1}, all
the Skyrme parameters including $x_1$, $x_2$ and $x_3$ have been
derived, listed in Table~\ref{Gcoupling}. Similar to the result
given in Ref.~\cite{qmcnpa1}, the obtained values of the $x_0$,
$t_0$, $t_1$, $t_2$, $t_3$ and $W_0$ parameters are close to the
SkM$^*$ force.  In Ref.~\cite{qmcnpa1}, the parameters $x_1$, $x_2$
and $x_3$ are set to be zero. However, it was pointed out that the
parameter $x_3$ has significant effect on the symmetry
energy~\cite{SHF18}. There already exist many sets of Skyrme
parameters. It is not our aim to derive a new set of parameters. In
the present work, we extend the understanding of the Skyrme force in
the framework of the QMC model, which has been done primarily in
Ref.~\cite{qmcnpa1}.

\begin{table}\centering\caption{The Skyrme parameters obtained in the present work (labeled by Sqmc) from the QMC model~\cite{qmcnpa1}. \label{Gcoupling}}
\vspace{1cm} \setlength{\tabcolsep}{20pt}
\begin{tabularx}{200pt}{lr}
\hline
Skyrme Parameters&\\
\hline
     $t_0$ (MeV fm$^3$) & $-2648.19$\\
 $t_1$ (MeV fm$^5$) & 371.07\\
   $t_2$ (MeV fm$^5$) & $-121.644$ \\
             $t_3$ (MeV fm$^4$) & 15553.495 \\
$x_0$ & 0.60146 \\
$x_1$ &0.2697 \\
 $x_2$ & $-0.23701$ \\
 $x_3$ & 0.6968 \\
 $W_0$ (MeV fm$^5$) & 104.498 \\
$\alpha$ & $1/6$\\
 \hline
\end{tabularx}

\end{table}

\section{Calculations and discussions}
\begin{figure}
\includegraphics[scale=0.45]{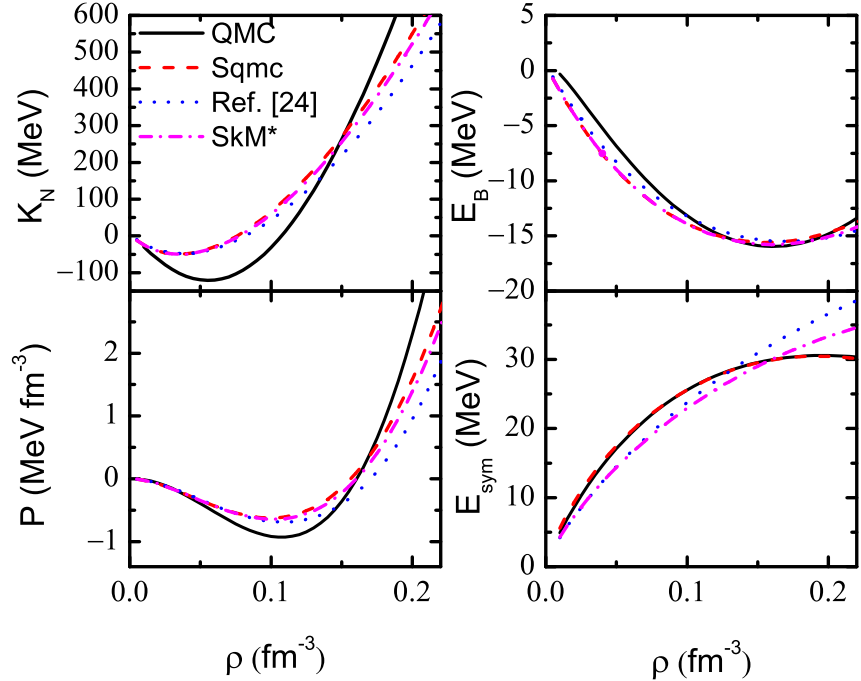}
\caption{(Color online) Comparisons between the QMC and Skyrme
calculations of the properties of the nuclear matter, i.e.,
incompressibility $K_{\rm N}$, pressure $P$, binding energy $E_{\rm
B}$ and symmetry energy $E_{\rm sym}$ within the Hartree-Fock
approximation. For the SHF calculations, SkM$^*$, Sqmc and the
Skyrme parameters given in Ref.~\cite{qmcnpa1} have been used.
}\label{nuclearmatterproperty}\end{figure}

As the first step, we compare the Hartree-Fock calculations of the
QMC and Skyrme models with different parameters for the nuclear
matter, shown in fig.~\ref{nuclearmatterproperty}. It is found that
the calculations give similar properties of the nuclear matter
around the saturation density. However, different behaviors are seen
in densities away from the saturation. The Sqmc parameters which
contain the spin-exchange parameters $x_1$, $x_2$ and $x_3$ give a
similar behavior of the symmetry energy against density as the QMC
model. The symmetry energy curve calculated by the Skyrme force
neglecting the $x_1$, $x_2$ and $x_3$ parameters~\cite{qmcnpa1} is
much more stiff than that obtained by the QMC model. A soft symmetry
energy against density seems more acceptable from the recent
analysis of experimental data~\cite{softsymmetrycurve}.

\begin{figure}
\hspace{-0.8  cm}\includegraphics[scale=0.5]{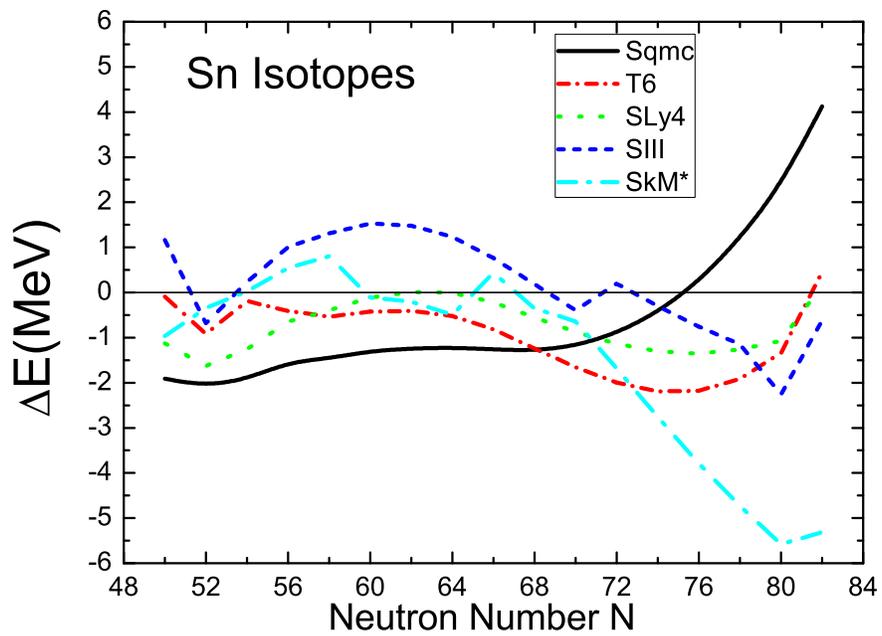}
\caption{(Color online) Difference between experimental and
calculated binding energies for even-even Sn isotopes. The
calculations are done using SHF+BCS with different sets of Skyrme
parameters.} \label{wang-fig-sn.eps}
\end{figure}

\begin{figure}
\hspace{-0.8  cm}\includegraphics[scale=0.5]{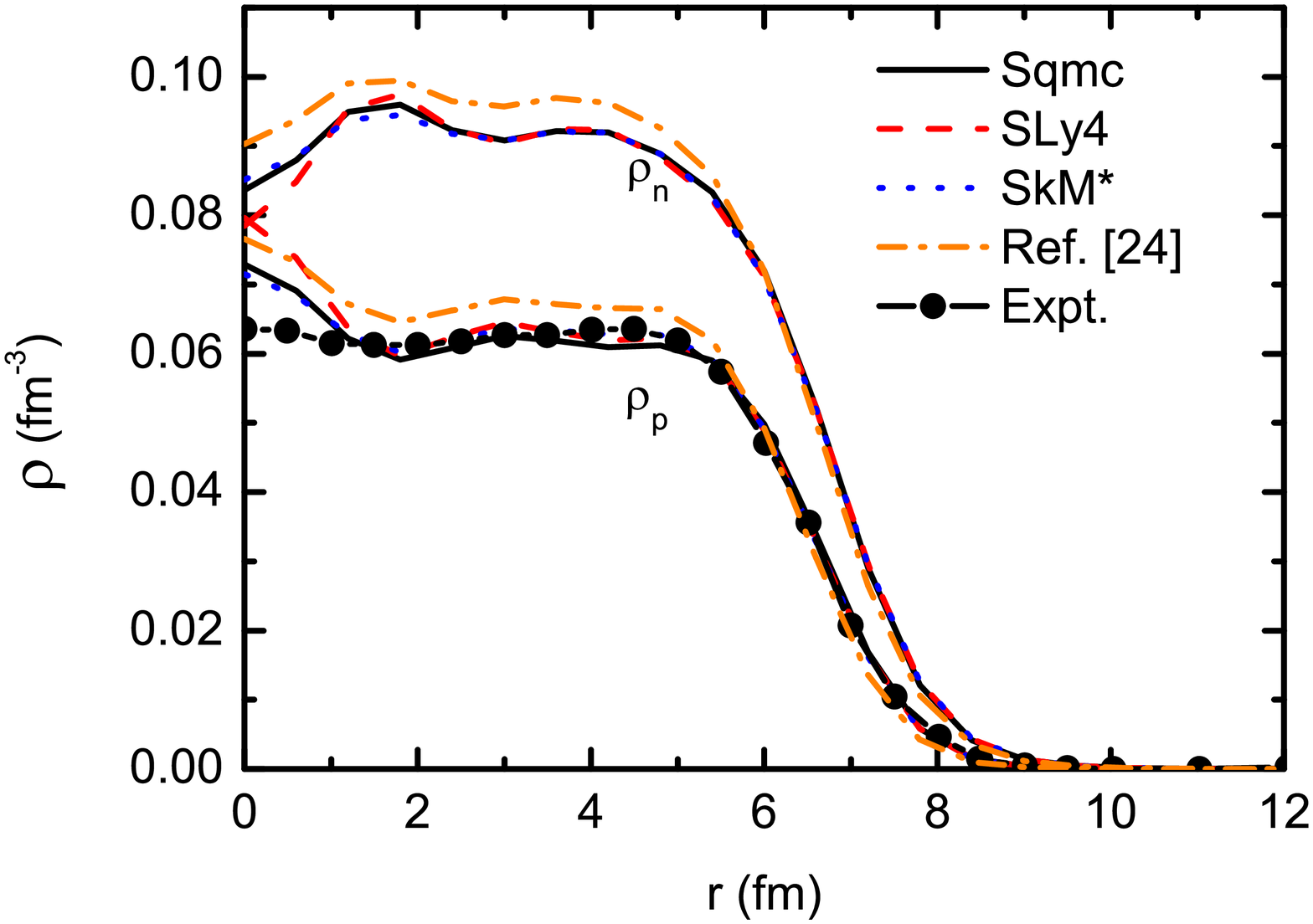}
\caption{(Color online) Neutron and proton density distributions for
$^{208}$Pb (indicated by $\rho_n$ and $\rho_p$, respectively)
calculated with Sqmc, SLy4, SkM$^*$ and the Skyrme parameters given
in Ref.~\cite{qmcnpa1}, compared with the experimental charge
density~\cite{chargedensity}.}\label{pbdensity}\end{figure}

As a further test, we calculate the properties of finite nuclei by
the Sqmc parameters. We use the SHF-BCS code of ev8~\cite{ev8} with
a self-consistent two-body center-of-mass correction. Calculations
for the double magic nuclei have shown a significant improvement,
compared with those in Ref.~\cite{qmcnpa1} where the parameters
$x_1$, $x_2$ and $x_3$ were set zero. We have also calculated
binding energies for the long chain of Sn isotopes from $^{100}$Sn
to $^{132}$Sn, for which the paring strength is fitted by
reproducing the neutron separation energies of Sn isotopes, shown in
Fig.~\ref{wang-fig-sn.eps}. We see that reasonable agreements
between data and calculations are obtained with the Sqmc parameters.
Here we should remind that the Sqmc parameters are obtained by
comparing the energy density functionals of the QMC and SHF models
for the nuclear matter without fitting the properties of finite
nuclei, while other sets of the Skyrme parameters were determined by
the fitting procedure to the experimental data of finite nuclei
(including the data of binding energies).

Fig.~\ref{pbdensity} shows the calculated neutron and proton density
distributions for $^{208}$Pb, compared with the experimental charge
density~\cite{chargedensity}. The Sqmc force gives a similar density
distribution to those obtained by the SLy4 and SkM$^*$ forces,
agreeing well with the data. We see that the calculations with the
inclusion of the spin-exchange terms are improved
noticeably.

In mean-field models, the concept of the effective mass of the
nucleon mainly describes effects
related to the nonlocality of the underlying nuclear interactions
and the Pauli exchange effects in nuclear
system~\cite{emassimportant}. The effective mass for the nuclear
matter in the SHF model can be written as~\cite{SHFeq}
\begin{eqnarray}\label{emshf}
\nonumber\frac{\hbar^2}{2M_f^{*}}\tau_f = &&\tau_f \left\{
\frac{\hbar^2}{2M}
+\frac{1}{8}\rho\left[t_1(2+x_1)+t_2(2+x_2)\right]
\right.\\&&\left.+\frac{1}{8}\rho_f
\left[t_2(1+2x_2)-t_1(1+2x_1)\right] \right\},
\end{eqnarray}
where the index ``$f$" indicates the proton or neutron. In the QMC
model, the non-relativistic effective mass can be extracted when one
collects the $\rho\tau$ terms,
\begin{eqnarray}\label{emqmchf}
\nonumber\frac{\hbar^2}{2M_f^{*}}\tau_f=&&\tau_f \left\{
\frac{\hbar^2}{2M}+\frac{1}{2}\frac{G_\sigma}{M^2}\frac{\rho}{1+dG_\sigma\rho}
+\frac{1}{4}\frac{G_\rho}{m_\rho^2}\rho\right.\\&&\left.+\rho_f
\left(-\frac{G_\rho}{8m_\rho^2}-\frac{G_\sigma}{2m_\sigma^2}+
\frac{G_\omega}{2m_\omega^2}-\frac{G_\sigma}{4M^2}\right)\right\}.
\end{eqnarray}
The first two terms are the Hartree contribution while the remaining
terms come from the Fock exchange.

\begin{figure}
\hspace{-0.8  cm}\includegraphics[scale=0.5]{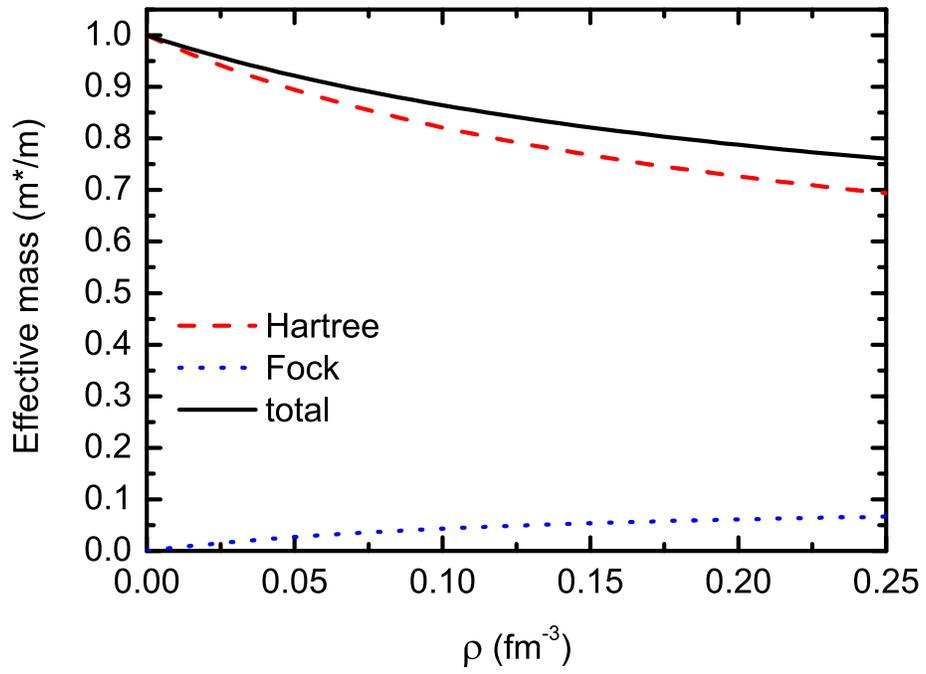}
\caption{(Color online) The density dependence of the
non-relativistic effective mass, $m^*/m$, calculated by the QMC
model for the symmetric nuclear matter.} \label{emhfvsh}
\end{figure}

In the SHF model, the effect from spin-exchange terms with the
parameters $x_1$, $x_2$ and $x_3$ appear explicitly in the effective
mass of Eq.~(\ref{emshf}). For the QMC model, if the effect from the
Fock exchange is neglected, the non-relativistic effective mass
contains only the isovector term. (The isovector effective mass can
be obtained by setting $\rho_f=0$ in Eq.~(\ref{emqmchf}).) If the
Fock term is taken into account, the $\rho$ meson has contribution
to both the isovector and isoscalar effective mass. The
non-relativistic effective mass calculated by the QMC model for the
symmetric nuclear matter is shown in Fig.~\ref{emhfvsh}.

\begin{figure}
\hspace{-0.8  cm}\includegraphics[scale=0.5]{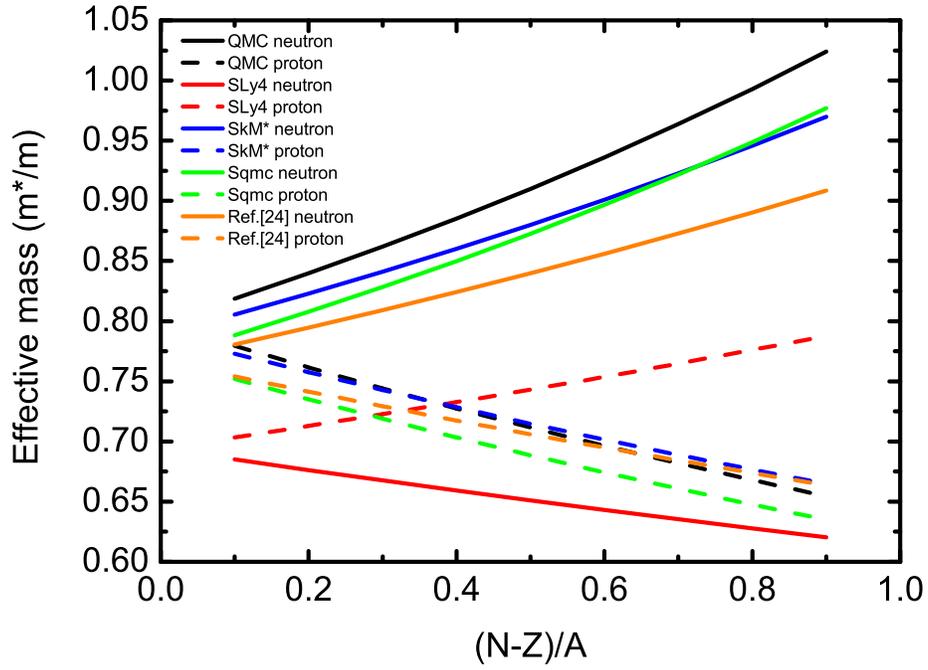}
\caption{(Color online) Calculated non-relativistic effective masses
($m^*/m$) of the proton and neutron against the isospin in the QMC
calculations, compared with the SHF calculations with Sqmc, SLy4,
SkM$^*$ and the Skyrme parameters given in Ref.~\cite{qmcnpa1}. The
calculations are done at the saturation density of $\rho=0.16$
fm$^{-3}$.}\label{emqmcshf}\end{figure}

The longstanding problem remains about the neutron-proton effective
mass splitting, which is crucial for extracting the density
dependence of symmetry energy from nuclear reaction using transport
models~\cite{Libaoan,transport2}. The dominant contribution to the
neutron-proton effective mass splitting is from the isovector
channel of the energy density functional. Fig.~\ref{emqmcshf} gives
the changes of the effective masses with the isospin, compared with
SHF calculations. The behavior of effective masses in the QMC
calculations is similar to that of the SkM$^*$ calculations. It
seems more acceptable recently that the effective mass of neutron is
bigger than that of proton~\cite{emassbaoan,emassdbhf}.

In Ref.~\cite{3bodytensor} it was shown that the three-body term has
an important influence on the symmetry energy curves. In
Fig.~\ref{nuclearmatterproperty}, we have already given the symmetry
energy for the nuclear matter, calculated by the QMC and SHF models
within the Hartree-Fock approximation. In order to gain a more
detailed insight into the nature of the symmetry energy and the
effect from the Fock exchange, we make further analysis. The QMC
symmetry energy for the nuclear matter within the Hartree
approximation is written as,
\begin{equation}\label{h symmetry energy}
a_{\rm sym}^{\rm H}=
\frac{\hbar^2}{6M}(\frac{3\pi^2}{2})^\frac{2}{3}\rho^\frac{2}3+
\frac{G_\sigma}{6M^2}(\frac{3\pi^2}2)^\frac{2}3
\frac{\rho^\frac{5}3}{1+dG_\sigma\rho}+\frac{G_\rho}{8}\rho.
\end{equation}
If the Fock exchange is taken into account, the QMC Hartree-Fock
symmetry energy for the nuclear matter becomes,
\begin{eqnarray}\label{qmchf symmetry energy}
\nonumber a_{\rm sym}^{\rm HF}&=&
\frac{\hbar^2}{6M}\left(\frac{3\pi^2}{2}\right)^\frac{2}{3}\rho^\frac{2}3+
\frac{1}3\left(\frac{3\pi^2}{2}\right)^\frac{2}3
\left(\frac{G_\omega}{m_\omega^2}-\frac{G_\sigma}{m_\sigma^2}\right)
\rho^\frac{5}3\\&+&\left[\frac{5G_\rho}{32}+\frac{G_\sigma}{8\left(1+d\rho
G_\sigma\right)^3}-\frac{G_\omega}{8}\right]\rho.
\end{eqnarray}
In the SHF model, the symmetry energy for the nuclear matter has the
following form~\cite{SHFeq},
\begin{eqnarray}\label{skhf symmetry energy}
\nonumber a_{\rm sym}^{\rm SHF}& =&\frac{\hbar^2}{6M}
\left(\frac{3\pi^2}{2}\right)^\frac{2}{3}
\rho^\frac{2}3-\frac{1}{24} \left(\frac{3\pi^2}2\right)^\frac{2}3
\left[3t_1x_1-t_2\left(4+5x_2\right)\right]\rho^\frac{5}3\\
&&-\frac{1}8t_0
\left(2x_0+1\right)\rho-\frac{1}{48}t_3\left(2x_3+1\right)\rho^{\alpha+1}.
\end{eqnarray}

\begin{figure}
\hspace{-0.8  cm}\includegraphics[scale=0.5]{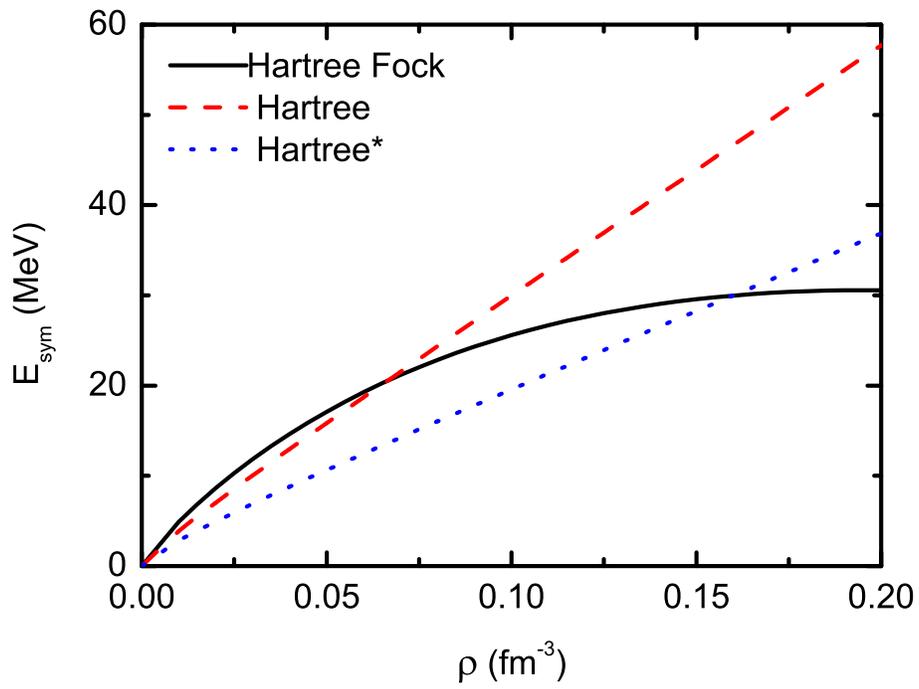}
\caption{(Color online) Symmetry energy curves calculated by the QMC
model in the Hartree and Hartree-Fock approximations. The dashed
line is obtained by the Hartree approximation using the same
parameter values as in the Hartree-Fock calculation. The dotted line
is also for the Hartree calculation but with parameters readjusted
by re-fitting nuclear matter properties at the saturation density.}
\label{ashfvsh}
\end{figure}

\begin{figure}
\hspace{-0.8  cm}\includegraphics[scale=0.5]{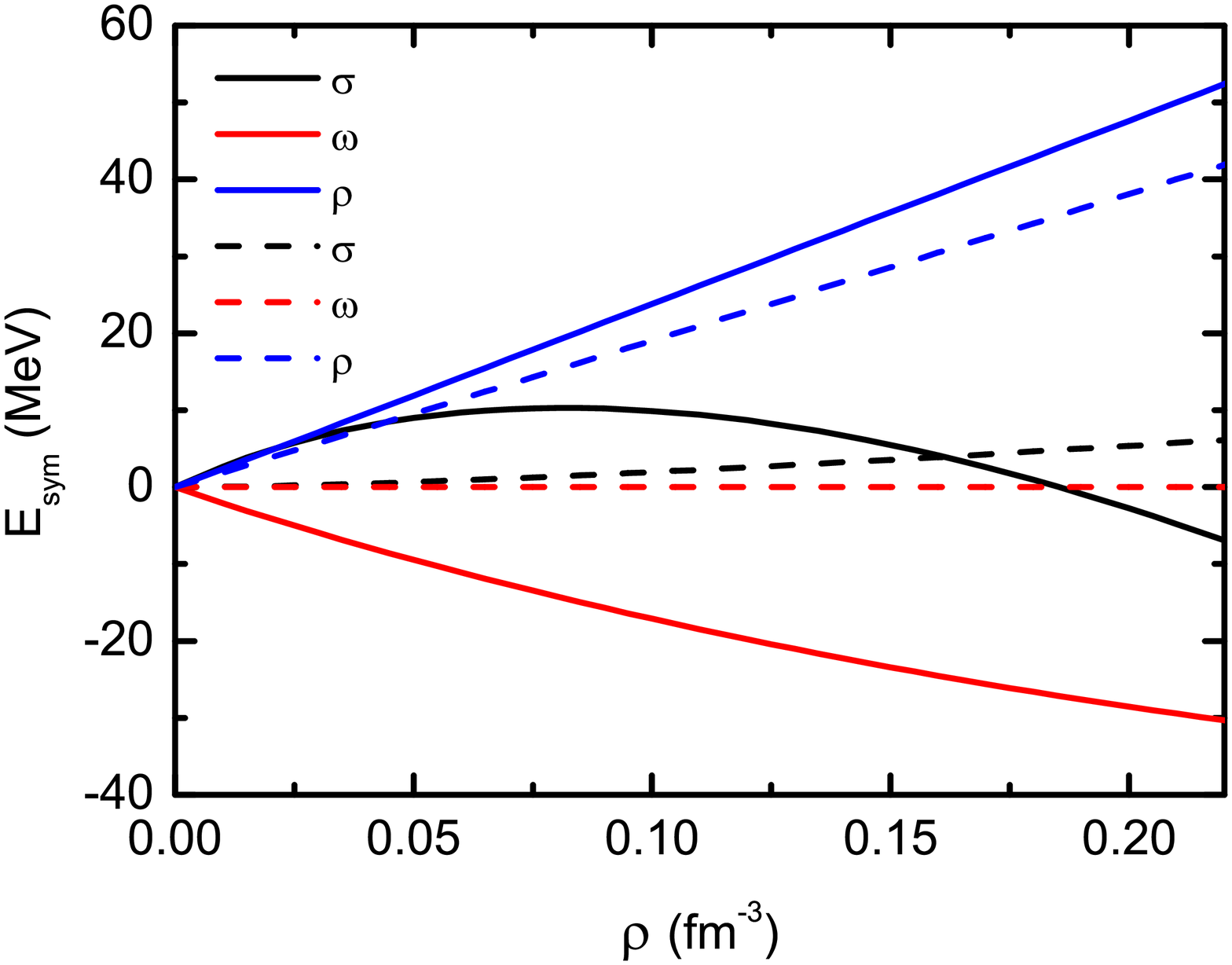}
\caption{(Color online) Contributions of the $\sigma$, $\omega$, and
$\rho$ mesons to the symmetry energy. The solid lines are for the
QMC Hartree-Fock calculations, and the dashed lines are for the QMC
Hartree calculations with the same parameters.}\label{ashfhmeson}
\end{figure}

The symmetry energy contains two contributions: 1) from the kinetic
energy; 2) from the interaction. It is seen that the QMC and SHF
models have similar density-dependent forms for the kinetic energy
part which originates from different Fermi momenta of nucleons and
the momentum dependence of the isoscalar potential. The momentum
dependence of the isoscalar potential has been well analyzed with
experimental data~\cite{momentum1,momentum2}. The different behavior
of the density dependence in the symmetry energy would be caused by
the interaction part which is from the isovector potential of the
energy density functional. From Eq.~(\ref{h symmetry energy}) in the
Hartree approximation, we see that the symmetry energy contributed
from the interaction part, $\frac{G_\rho}{8}\rho$, has a linear form
respecting the density which dominates the symmetry energy. The Fock
exchange leads to a nonlinear density-dependent interaction part for
the symmetry energy, see Eq.~(\ref{qmchf symmetry energy}) for the
QMC model. Fig.~\ref{ashfvsh} displays the curves of the symmetry
energy with and without the Fock exchange. It is seen that the
inclusion of the Fock exchange gives a strong density-dependent
symmetry energy. In the figure, we have also given the symmetry
energy curve of the Hartree calculation with re-fitted parameters
(readjust the QMC coupling constants by fitting the properties of
the nuclear matter at the saturation) to see if the effect of the
Fock exchange can be included by adjusting parameters. We see that
the effect from the Fock exchange can not be included by readjusting
parameters. It has been pointed out that models within the Hartree
approximation give nearly-linear symmetry
energy~\cite{Libaoansymmetry}.

It is seen from Eq.~(\ref{qmchf symmetry energy}) that the $\sigma$
meson brings a density dependent contribution to the interaction
part of the symmetry energy through the Fock exchange. In
Fig.~\ref{ashfhmeson}, we show the contributions of the mesons to
the symmetry energy in the QMC-Hartree and QMC-Hartree-Fock models.
We see that, with the Fock exchange included, the contributions of
the $\rho$ and $\omega$ mesons have nearly linear but opposite
trends against the density. These two parts can be canceled partly.
The $\sigma$ meson through the Fock exchange produces a strong
density dependence, which increases first with density and then
decreases drastically in higher density, leading to a soft behavior
of the symmetry energy.

\begin{figure}
\hspace{-0.8  cm}\includegraphics[scale=0.5]{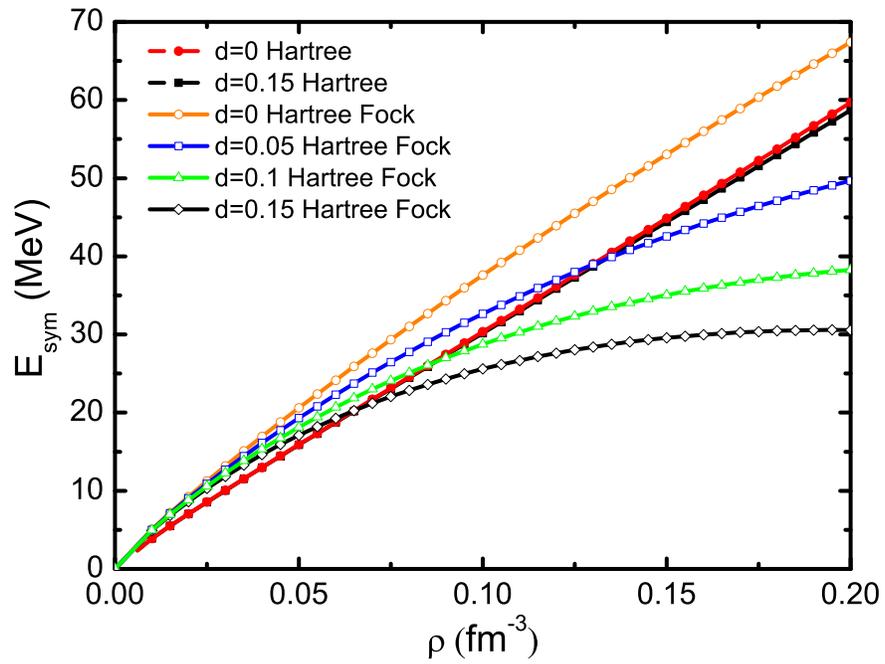}
\caption{(Color online) Symmetry energy curves calculated by the QMC
model in the Hartree and Hartree-Fock approximations with assuming
different values of the parameter $d$. The dashed lines are obtained
by the Hartree approximation and the solid lines are obtained by the
Hartree-Fock calculation. All the calculations use the same
parameters as QMC Hartree-Fock except the parameter $d$.}
\label{dsymmetry}
\end{figure}

We further study effects on the symmetry energy from the
density-dependent meson-nucleon exchange interaction. If the QMC
model is compared with RMF model, one would find that the main
difference is from the scalar field~\cite{QMCGreiner}. The scalar
field in the QMC model has a nonlinear density dependence, which is
derived by considering subnucleonic
freedoms~\cite{qmcnpa1,QMCreview,QMCGreiner}. This nonlinear
behavior originates from the scalar polarizability parameter $d$ of
the nucleon. In the MIT bag model, it can be well represented by the
bag radius~\cite{qmcnpa1,qmcnpa2,qmcnpa3}. Fig.~\ref{dsymmetry}
plots the density dependence of the symmetry energy with assuming
the different values of the parameter $d$. The standard value of the
parameter $d$ is $0.150$ fm which corresponds to a bag radius of
$0.8$ fm~\cite{qmcnpa1}. If we set $d=0$ which is equivalent to
assuming a point particle for the nucleons, the meson exchange
interaction becomes density independent, which gives approximately
linear symmetry energy, see Fig.~\ref{dsymmetry}. With increasing
the parameter $d$, the symmetry energy becomes soft against the
density. However, the Hartree calculation still give stiff linear
symmetry energy. By comparing these calculations, we can conclude
that the nonlinear scalar field can soften the symmetry energy curve
through the Fock exchange interaction.
As mentioned already in the introduction of the paper, the symmetry
energy is still a debating problem. Different models give quite
different predictions of the symmetry
energy~\cite{stone,Libaoan,neutronskin,softsymmetrycurve,Vmbsymmetry1,Vmbsymmetry2,Vmbsymmetry3,dbhfsymmetry,Libaoansymmetry,Liu02,bhfsymmetry}.
However, a soft symmetry energy seems to have been supported by the
recent analysis of experimental data~\cite{softsymmetrycurve}.

\section{Summary and conclusion}
With the inclusion of spin-exchange in the Skyrme force, we have
discussed the relation between the Skyrme and meson-nucleon exchange
forces. As in Ref.~\cite{qmcnpa1}, the Skyrme force can be
understood within the QMC model. By comparing the energy density
functionals, a set of Skyrme parameters including the Skyrme
spin-exchange parameters $x_1$, $x_2$ and $x_3$ is obtained, which
can give reasonable results for the properties of nuclear matter and
finite nuclei. However, the determination of the parameters is not
the aim of the paper. Many sets of the parameters for the Skyrme
force have existed with fitting to the experimental data of finite
nuclei also.

We have paid special attention to the isovector channel of the
nuclear energy density functionals. The QMC and SHF models have a
similar density dependence for the isoscalar channel, but have
different forms for the isovector channel. It is found that the Fock
exchange leads to a nonlinear density-dependent isovector channel in
the QMC model, which makes a soft symmetry energy. The nonlinear
density dependence is produced by the scalar field which is derived
by considering subnucleonic freedoms.

\section{Acknowledgement}

This work has been supported by the Chinese Major State Basic
Research Development Program under Grant 2007CB815000 and the
National Natural Science Foundation of China under Grant Nos.
10735010 and 10975006. C.Q. acknowledges the support from the
Swedish Research Council (VR) under grant
Nos. 623-2009-7340 and 2010-4723.

\bibliographystyle{elsarticle-num}

\end{document}